\documentclass[aps,prd,reprint,floatfix,superscriptaddress,lengthcheck,sort&compress,letterpaper,nofootinbib]{revtex4-1}

\usepackage{graphicx}
\usepackage{dcolumn}
\usepackage{bm}
\usepackage{dsfont}
\usepackage{amsmath}
\usepackage[usenames]{color}
\usepackage[colorlinks=true,linkcolor=blue,citecolor=blue,urlcolor=blue]{hyperref}

      \def\longlonglongrightarrow{
      \relbar\joinrel\relbar\joinrel\relbar\joinrel\relbar\joinrel\relbar\joinrel\relbar\joinrel\rightarrow}

\begin{document}

\title{Reply to comment on "The $\sigma$-meson: four-quark vs. two-quark components and decay width in a Bethe-Salpeter approach"}

\author{Nico Santowsky}
 \email[e-mail: ]{nico.santowsky@theo.physik.uni-giessen.de}
\affiliation{Institut f\"ur Theoretische Physik, Justus-Liebig Universit\"at Gie{\ss}en, 35392 Gie{\ss}en, Germany}
\author{Gernot Eichmann}
\email[e-mail: ]{gernot.eichmann@tecnico.ulisboa.pt}
\affiliation{LIP Lisboa, Av.~Prof.~Gama~Pinto 2, 1649-003 Lisboa, Portugal}
\affiliation{Departamento de F\'isica, Instituto Superior T\'ecnico, 1049-001 Lisboa, Portugal}
\author{Christian S. Fischer}
 \email[e-mail: ]{christian.fischer@theo.physik.uni-giessen.de}
\affiliation{Institut f\"ur Theoretische Physik, Justus-Liebig Universit\"at Gie{\ss}en, 35392 Gie{\ss}en, Germany}
\affiliation{Helmholtz Forschungsakademie Hessen f\"ur FAIR (HFHF),
	GSI Helmholtzzentrum f\"ur Schwerionenforschung, Campus Gie{\ss}en, 35392 Gie{\ss}en, Germany}
\author{Paul C. Wallbott}
\affiliation{Institut f\"ur Theoretische Physik, Justus-Liebig Universit\"at Gie{\ss}en, 35392 Gie{\ss}en, Germany}
\author{Richard Williams}
\affiliation{Institut f\"ur Theoretische Physik, Justus-Liebig Universit\"at Gie{\ss}en, 35392 Gie{\ss}en, Germany}

\date{\today}

\begin{abstract}
In a recent comment \cite{Blankleider:2021odu} Blankleider and Kvinikhidze claim that our work \cite{Santowsky:2020pwd} is based on a set of
inconsistent Bethe-Salpeter equations for the coupling of four-quark to two-quark states. Here we demonstrate that their argument is insubstantial.
\end{abstract}

\maketitle

In continuum QCD, four-quark ($qq\bar{q}\bar{q}$) states are described by four-body Faddeev-Yakubovsky equations derived from the quark eight-point function.
The details of the corresponding interaction kernels have been worked out in \cite{Huang:1974cd,Khvedelidze:1991qb,Yokojima:1993np} and put to
work in a series of publications in the past years \cite{Heupel:2012ua,Eichmann:2015cra,Eichmann:2015nra,Wallbott:2019dng,Wallbott:2020jzh,Santowsky:2020pwd},
see \cite{Eichmann:2020oqt} for an overview.
Applications include studies of the light meson sector as well as the charmonium energy region, where a number of experimentally discovered
'XYZ states' may very well feature substantial four-quark contributions \cite{Esposito:2016noz,Guo:2017jvc,Ali:2017jda,Liu:2019zoy,Brambilla:2019esw}.

One of the many interesting features of four-quark states is their potential mixing with two-quark ($q\bar{q}$) states of the same quantum numbers. This is especially relevant in isospin singlet channels. In our recent work \cite{Santowsky:2020pwd} we therefore put forward a phenomenologically
motivated scheme to provide for the coupling of the four-body $qq\bar{q}\bar{q}$ equations to two-body $q\bar{q}$ equations starting from an ansatz
for the T-matrix. We discussed 
the merits and shortcomings of our scheme and pointed out explicitly that a more formal derivation is called for in future 
work \cite{Santowsky:2020pwd}. Nevertheless we believed, and still believe, that this scheme is suitable for a first study of the mixing effects
and can be expected to deliver meaningful results.

This scheme has been criticized by Blankleider and Kvinikhidze in a recent comment \cite{Blankleider:2021odu}. In order to be able 
to appreciate their argument we briefly recapitulate some basics notions of Bethe-Salpeter equations (BSEs). In general, 
homogeneous BSEs are derived from the $(2n)$-(anti-)quark scattering matrix $T^{(n)}$ and the corresponding
scattering kernel $K^{(n)}$:
\begin{equation}
	T^{(n)}=K^{(n)}+K^{(n)}G_0^{(n)}T^{(n)}\,.\label{eq-dysonsum}
\end{equation}
In this equation, $G_0^{(n)}$ denotes the product of $n$ dressed (anti-)quark propagators. One then defines a bound-state amplitude (BSA)
$\Gamma^{(n)}$ at the physical pole $Q^2 \rightarrow -M^2$ of the $T$ matrix via
\begin{equation}
	T^{(n)} \overset{Q^2 \rightarrow -M^2}{\longlonglongrightarrow}  \frac{\Gamma\bar\Gamma}{Q^2+M^2}\,.\label{eq-bsaansatz}
\end{equation}
Inserting this expression into Eq.~(\ref{eq-dysonsum}) leads to the homogeneous on-shell $n$-quark BSE:
\begin{equation}
	\Gamma^{(n)}=K^{(n)}G_0^{(n)}\Gamma^{(n)} \,.\label{eq-bsegeneral}
\end{equation}
\begin{figure*}[t]
	\centering
	\includegraphics[scale=1.00]{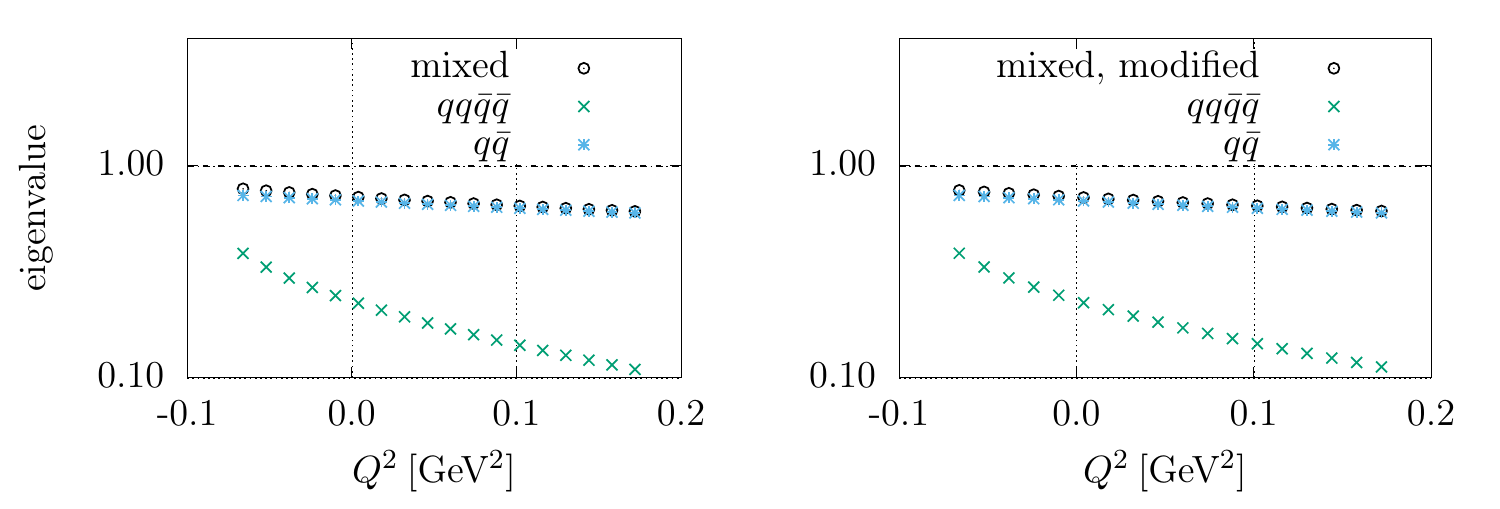}
	\caption{The eigenvalue curves $\lambda(Q^2)$ for calculations with different active components in the Bethe-Salpeter wave functions,
		i.e., $q\bar{q}$ only, $qq\bar{q}\bar{q}$ only, and the curve for the coupled system of equations ('mixed'), see \cite{Santowsky:2020pwd}
	    for details. Bound states/resonances occur at $Q^2=-M^2$.
		The diagram on the left shows the results in the setup of Ref.~\cite{Santowsky:2020pwd}, which may be compared to the
		results using the modified kernel, Eq.~(\ref{mod}), displayed in the diagram on the right.
		\label{fig-evcurves}}
\end{figure*}
In our work, we derived such a kernel for the BSE of a quark-antiquark bound state that includes the coupling to four-quark states. In
the notation of Ref.~\cite{Blankleider:2021odu}, the resulting BSE for the quark-antiquark BSA $\Gamma^*$ can be written as:
\begin{equation}\label{Eq9}
\Gamma^* = \left[ K^{(2)} + K^{(2)} G_0^{(2)} \Omega \right]  G_0^{(2)} \Gamma^* \,,
\end{equation}
where $\Omega= \bar{N} G^0 (1-VG^0)^{-1} N = \bar{N} G N$ 
depends on the Green function $G$ of the four-body system without coupling to $q\bar{q}$ states,
the $\bar{N},N$ describe transitions between two- and four-body states, $G^0$ denotes propagation and $V$ interaction
in the four-body state. The details of $\Omega$ do not matter for the argument.

In their comment \cite{Blankleider:2021odu}, Blankleider and Kvinikhidze note (correctly) that the kernel in Eq.(\ref{Eq9}) 
is $q\bar{q}$-reducible due to the presence of the $q\bar{q}$ propagator $G_0^{(2)}$ between $K^{(2)}$ and $\Omega$. 
From this observation they conclude that the equation therefore must be inconsistent. 

In the following we show that this notion is misguided. 
Consider again the defining equation for the q$\bar{q}$-irreducible kernel $K^{(2)}$ via the T-matrix $T^{(2)}$ and a product of propagators $G_0^{(2)}$:
\begin{align}
	T^{(2)} & = K^{(2)}+K^{(2)}G_0^{(2)}T^{(2)}  \nonumber\\
	        & = K^{(2)}+K^{(2)}G_0^{(2)}K^{(2)} + K^{(2)}G_0^{(2)}K^{(2)}G_0^{(2)}T^{(2)} \nonumber \\
            & = \dots
\end{align}
Here, the second line has been obtained by once inserting $T^{(2)}$ on the right hand side of the first line. Thus, both lines (and all further iterations) are exact.
Homogenous BSEs can be extracted from either line by inserting (\ref{eq-bsaansatz}) and taking the
limit $Q^2 \rightarrow -M^2$. One obtains different forms of BSEs:
\begin{align}
\Gamma^{(2)} &= K^{(2)}G_0^{(2)}\Gamma^{(2)}  \nonumber\\
             &=	K^{(2)}G_0^{(2)} K^{(2)}G_0^{(2)}\Gamma^{(2)} \nonumber \\
             & = \dots
\end{align}
which must have identical solutions because all of them have been derived from exact equations for the T-matrix. However, whereas the
kernel $K^{(2)}$ of the first BSE is by definition $q\bar{q}$-irreducible, the composite kernel $K^{(2)}G_0^{(2)} K^{(2)}$ of the second
BSE (and higher iterations) does not have this property (note the presence of the propagator $G_0^{(2)}$).

This little example shows that the appearance of reducible kernels in BSEs is certainly not a sufficient signal of inconsistency.
Thus, the argument presented in the comment \cite{Blankleider:2021odu} is not correct.

For the sake of the argument, however, let us play along and consider the implications of replacing the kernel in Eq.~(\ref{Eq9})
with
\begin{equation}\label{mod}
	\Gamma^* = \left[ K^{(2)} + \Omega  \right] G_0^{(2)} \Gamma^* \,.
\end{equation}
Here, we have removed the 'offending part' $K^{(2)} G_0^{(2)}$ from the second term. 
The resulting kernel is $q\bar{q}$-irreducible and identical to the one suggested by Kvinikhidze and Blankleider
in a very recent work \cite{Kvinikhidze:2021kzu}. We repeated the calculation of Ref.~\cite{Santowsky:2020pwd} on the mixing between
the four-quark and the two-quark state for the light isospin-singlet $0^{++}$ meson. Our results for the eigenvalue curves $\lambda(Q^2)$
of the meson are shown in Fig.~\ref{fig-evcurves}, where we compare results for the kernel of Eq.~(\ref{Eq9}) 
used in Ref.~\cite{Santowsky:2020pwd} (left diagram)
to results using the kernel of Eq.~(\ref{mod}) (right diagram). The deviations between the two setups are well below the 5\% level and 
of the same order as the numerical error of our calculation. The extrapolated meson mass with modified kernel, Eq.~(\ref{mod}),
is given by
\begin{equation}
	M_{0^{++}} = 484\,\pm\,37 \,\mbox{MeV}\,.
\end{equation}
Compared to the result $M_{0^{++}} = 456\,\pm\,24 \,\mbox{MeV}$ from Ref.~\cite{Santowsky:2020pwd}, we find agreement within
numerical accuracy and extrapolation error. The mean values are separated on the 5\% level. In any case, the change
of kernels does not materially alter our findings, and all conclusions discussed in Ref.~\cite{Santowsky:2020pwd} also apply to
the setup using Eq.~(\ref{mod}).


\bibliographystyle{apsrev4-1}
\bibliography{sigma_mixing}

\begin{thebibliography}{17}%
\makeatletter
\providecommand \@ifxundefined [1]{%
 \@ifx{#1\undefined}
}%
\providecommand \@ifnum [1]{%
 \ifnum #1\expandafter \@firstoftwo
 \else \expandafter \@secondoftwo
 \fi
}%
\providecommand \@ifx [1]{%
 \ifx #1\expandafter \@firstoftwo
 \else \expandafter \@secondoftwo
 \fi
}%
\providecommand \natexlab [1]{#1}%
\providecommand \enquote  [1]{``#1''}%
\providecommand \bibnamefont  [1]{#1}%
\providecommand \bibfnamefont [1]{#1}%
\providecommand \citenamefont [1]{#1}%
\providecommand \href@noop [0]{\@secondoftwo}%
\providecommand \href [0]{\begingroup \@sanitize@url \@href}%
\providecommand \@href[1]{\@@startlink{#1}\@@href}%
\providecommand \@@href[1]{\endgroup#1\@@endlink}%
\providecommand \@sanitize@url [0]{\catcode `\\12\catcode `\$12\catcode
  `\&12\catcode `\#12\catcode `\^12\catcode `\_12\catcode `\%12\relax}%
\providecommand \@@startlink[1]{}%
\providecommand \@@endlink[0]{}%
\providecommand \url  [0]{\begingroup\@sanitize@url \@url }%
\providecommand \@url [1]{\endgroup\@href {#1}{\urlprefix }}%
\providecommand \urlprefix  [0]{URL }%
\providecommand \Eprint [0]{\href }%
\providecommand \doibase [0]{http://dx.doi.org/}%
\providecommand \selectlanguage [0]{\@gobble}%
\providecommand \bibinfo  [0]{\@secondoftwo}%
\providecommand \bibfield  [0]{\@secondoftwo}%
\providecommand \translation [1]{[#1]}%
\providecommand \BibitemOpen [0]{}%
\providecommand \bibitemStop [0]{}%
\providecommand \bibitemNoStop [0]{.\EOS\space}%
\providecommand \EOS [0]{\spacefactor3000\relax}%
\providecommand \BibitemShut  [1]{\csname bibitem#1\endcsname}%
\let\auto@bib@innerbib\@empty
\bibitem [{\citenamefont {Blankleider}\ and\ \citenamefont
  {Kvinikhidze}(2021)}]{Blankleider:2021odu}%
  \BibitemOpen
  \bibfield  {author} {\bibinfo {author} {\bibfnamefont {B.}~\bibnamefont
  {Blankleider}}\ and\ \bibinfo {author} {\bibfnamefont {A.~N.}\ \bibnamefont
  {Kvinikhidze}},\ }\href@noop {} {\  (\bibinfo {year} {2021})},\ \Eprint
  {http://arxiv.org/abs/2102.05818} {arXiv:2102.05818 [hep-ph]} \BibitemShut
  {NoStop}%
\bibitem [{\citenamefont {Santowsky}\ \emph {et~al.}(2020)\citenamefont
  {Santowsky}, \citenamefont {Eichmann}, \citenamefont {Fischer}, \citenamefont
  {Wallbott},\ and\ \citenamefont {Williams}}]{Santowsky:2020pwd}%
  \BibitemOpen
  \bibfield  {author} {\bibinfo {author} {\bibfnamefont {N.}~\bibnamefont
  {Santowsky}}, \bibinfo {author} {\bibfnamefont {G.}~\bibnamefont {Eichmann}},
  \bibinfo {author} {\bibfnamefont {C.~S.}\ \bibnamefont {Fischer}}, \bibinfo
  {author} {\bibfnamefont {P.~C.}\ \bibnamefont {Wallbott}}, \ and\ \bibinfo
  {author} {\bibfnamefont {R.}~\bibnamefont {Williams}},\ }\href {\doibase
  10.1103/PhysRevD.102.056014} {\bibfield  {journal} {\bibinfo  {journal}
  {Phys. Rev. D}\ }\textbf {\bibinfo {volume} {102}},\ \bibinfo {pages}
  {056014} (\bibinfo {year} {2020})},\ \Eprint
  {http://arxiv.org/abs/2007.06495} {arXiv:2007.06495 [hep-ph]} \BibitemShut
  {NoStop}%
\bibitem [{\citenamefont {Huang}\ and\ \citenamefont
  {Weldon}(1975)}]{Huang:1974cd}%
  \BibitemOpen
  \bibfield  {author} {\bibinfo {author} {\bibfnamefont {K.}~\bibnamefont
  {Huang}}\ and\ \bibinfo {author} {\bibfnamefont {H.~A.}\ \bibnamefont
  {Weldon}},\ }\href {\doibase 10.1103/PhysRevD.11.257} {\bibfield  {journal}
  {\bibinfo  {journal} {Phys. Rev.}\ }\textbf {\bibinfo {volume} {D11}},\
  \bibinfo {pages} {257} (\bibinfo {year} {1975})}\BibitemShut {NoStop}%
\bibitem [{\citenamefont {Khvedelidze}\ and\ \citenamefont
  {Kvinikhidze}(1992)}]{Khvedelidze:1991qb}%
  \BibitemOpen
  \bibfield  {author} {\bibinfo {author} {\bibfnamefont {A.~M.}\ \bibnamefont
  {Khvedelidze}}\ and\ \bibinfo {author} {\bibfnamefont {A.~N.}\ \bibnamefont
  {Kvinikhidze}},\ }\href {\doibase 10.1007/BF01018820} {\bibfield  {journal}
  {\bibinfo  {journal} {Theor. Math. Phys.}\ }\textbf {\bibinfo {volume}
  {90}},\ \bibinfo {pages} {62} (\bibinfo {year} {1992})}\BibitemShut {NoStop}%
\bibitem [{\citenamefont {Yokojima}\ \emph {et~al.}(1993)\citenamefont
  {Yokojima}, \citenamefont {Komachiya},\ and\ \citenamefont
  {Fukuda}}]{Yokojima:1993np}%
  \BibitemOpen
  \bibfield  {author} {\bibinfo {author} {\bibfnamefont {S.}~\bibnamefont
  {Yokojima}}, \bibinfo {author} {\bibfnamefont {M.}~\bibnamefont {Komachiya}},
  \ and\ \bibinfo {author} {\bibfnamefont {R.}~\bibnamefont {Fukuda}},\ }\href
  {\doibase 10.1016/0550-3213(93)90459-3} {\bibfield  {journal} {\bibinfo
  {journal} {Nucl. Phys. B}\ }\textbf {\bibinfo {volume} {390}},\ \bibinfo
  {pages} {319} (\bibinfo {year} {1993})}\BibitemShut {NoStop}%
\bibitem [{\citenamefont {Heupel}\ \emph {et~al.}(2012)\citenamefont {Heupel},
  \citenamefont {Eichmann},\ and\ \citenamefont {Fischer}}]{Heupel:2012ua}%
  \BibitemOpen
  \bibfield  {author} {\bibinfo {author} {\bibfnamefont {W.}~\bibnamefont
  {Heupel}}, \bibinfo {author} {\bibfnamefont {G.}~\bibnamefont {Eichmann}}, \
  and\ \bibinfo {author} {\bibfnamefont {C.~S.}\ \bibnamefont {Fischer}},\
  }\href {\doibase 10.1016/j.physletb.2012.11.009} {\bibfield  {journal}
  {\bibinfo  {journal} {Phys. Lett.}\ }\textbf {\bibinfo {volume} {B718}},\
  \bibinfo {pages} {545} (\bibinfo {year} {2012})},\ \Eprint
  {http://arxiv.org/abs/1206.5129} {arXiv:1206.5129 [hep-ph]} \BibitemShut
  {NoStop}%
\bibitem [{\citenamefont {Eichmann}\ \emph {et~al.}(2016)\citenamefont
  {Eichmann}, \citenamefont {Fischer},\ and\ \citenamefont
  {Heupel}}]{Eichmann:2015cra}%
  \BibitemOpen
  \bibfield  {author} {\bibinfo {author} {\bibfnamefont {G.}~\bibnamefont
  {Eichmann}}, \bibinfo {author} {\bibfnamefont {C.~S.}\ \bibnamefont
  {Fischer}}, \ and\ \bibinfo {author} {\bibfnamefont {W.}~\bibnamefont
  {Heupel}},\ }\href {\doibase 10.1016/j.physletb.2015.12.036} {\bibfield
  {journal} {\bibinfo  {journal} {Phys. Lett.}\ }\textbf {\bibinfo {volume}
  {B753}},\ \bibinfo {pages} {282} (\bibinfo {year} {2016})},\ \Eprint
  {http://arxiv.org/abs/1508.07178} {arXiv:1508.07178 [hep-ph]} \BibitemShut
  {NoStop}%
\bibitem [{\citenamefont {Eichmann}\ \emph {et~al.}(2015)\citenamefont
  {Eichmann}, \citenamefont {Fischer},\ and\ \citenamefont
  {Heupel}}]{Eichmann:2015nra}%
  \BibitemOpen
  \bibfield  {author} {\bibinfo {author} {\bibfnamefont {G.}~\bibnamefont
  {Eichmann}}, \bibinfo {author} {\bibfnamefont {C.~S.}\ \bibnamefont
  {Fischer}}, \ and\ \bibinfo {author} {\bibfnamefont {W.}~\bibnamefont
  {Heupel}},\ }\href {\doibase 10.1103/PhysRevD.92.056006} {\bibfield
  {journal} {\bibinfo  {journal} {Phys. Rev.}\ }\textbf {\bibinfo {volume}
  {D92}},\ \bibinfo {pages} {056006} (\bibinfo {year} {2015})},\ \Eprint
  {http://arxiv.org/abs/1505.06336} {arXiv:1505.06336 [hep-ph]} \BibitemShut
  {NoStop}%
\bibitem [{\citenamefont {Wallbott}\ \emph {et~al.}(2019)\citenamefont
  {Wallbott}, \citenamefont {Eichmann},\ and\ \citenamefont
  {Fischer}}]{Wallbott:2019dng}%
  \BibitemOpen
  \bibfield  {author} {\bibinfo {author} {\bibfnamefont {P.~C.}\ \bibnamefont
  {Wallbott}}, \bibinfo {author} {\bibfnamefont {G.}~\bibnamefont {Eichmann}},
  \ and\ \bibinfo {author} {\bibfnamefont {C.~S.}\ \bibnamefont {Fischer}},\
  }\href {\doibase 10.1103/PhysRevD.100.014033} {\bibfield  {journal} {\bibinfo
   {journal} {Phys. Rev.}\ }\textbf {\bibinfo {volume} {D100}},\ \bibinfo
  {pages} {014033} (\bibinfo {year} {2019})},\ \Eprint
  {http://arxiv.org/abs/1905.02615} {arXiv:1905.02615 [hep-ph]} \BibitemShut
  {NoStop}%
\bibitem [{\citenamefont {Wallbott}\ \emph {et~al.}(2020)\citenamefont
  {Wallbott}, \citenamefont {Eichmann},\ and\ \citenamefont
  {Fischer}}]{Wallbott:2020jzh}%
  \BibitemOpen
  \bibfield  {author} {\bibinfo {author} {\bibfnamefont {P.~C.}\ \bibnamefont
  {Wallbott}}, \bibinfo {author} {\bibfnamefont {G.}~\bibnamefont {Eichmann}},
  \ and\ \bibinfo {author} {\bibfnamefont {C.~S.}\ \bibnamefont {Fischer}},\
  }\href@noop {} {\  (\bibinfo {year} {2020})},\ \Eprint
  {http://arxiv.org/abs/2003.12407} {arXiv:2003.12407 [hep-ph]} \BibitemShut
  {NoStop}%
\bibitem [{\citenamefont {Eichmann}\ \emph {et~al.}(2020)\citenamefont
  {Eichmann}, \citenamefont {Fischer}, \citenamefont {Heupel}, \citenamefont
  {Santowsky},\ and\ \citenamefont {Wallbott}}]{Eichmann:2020oqt}%
  \BibitemOpen
  \bibfield  {author} {\bibinfo {author} {\bibfnamefont {G.}~\bibnamefont
  {Eichmann}}, \bibinfo {author} {\bibfnamefont {C.~S.}\ \bibnamefont
  {Fischer}}, \bibinfo {author} {\bibfnamefont {W.}~\bibnamefont {Heupel}},
  \bibinfo {author} {\bibfnamefont {N.}~\bibnamefont {Santowsky}}, \ and\
  \bibinfo {author} {\bibfnamefont {P.~C.}\ \bibnamefont {Wallbott}},\ }\href
  {\doibase 10.1007/s00601-020-01571-3} {\bibfield  {journal} {\bibinfo
  {journal} {Few Body Syst.}\ }\textbf {\bibinfo {volume} {61}},\ \bibinfo
  {pages} {38} (\bibinfo {year} {2020})},\ \Eprint
  {http://arxiv.org/abs/2008.10240} {arXiv:2008.10240 [hep-ph]} \BibitemShut
  {NoStop}%
\bibitem [{\citenamefont {Esposito}\ \emph {et~al.}(2016)\citenamefont
  {Esposito}, \citenamefont {Pilloni},\ and\ \citenamefont
  {Polosa}}]{Esposito:2016noz}%
  \BibitemOpen
  \bibfield  {author} {\bibinfo {author} {\bibfnamefont {A.}~\bibnamefont
  {Esposito}}, \bibinfo {author} {\bibfnamefont {A.}~\bibnamefont {Pilloni}}, \
  and\ \bibinfo {author} {\bibfnamefont {A.~D.}\ \bibnamefont {Polosa}},\
  }\href {\doibase 10.1016/j.physrep.2016.11.002} {\bibfield  {journal}
  {\bibinfo  {journal} {Phys. Rept.}\ }\textbf {\bibinfo {volume} {668}},\
  \bibinfo {pages} {1} (\bibinfo {year} {2016})},\ \Eprint
  {http://arxiv.org/abs/1611.07920} {arXiv:1611.07920 [hep-ph]} \BibitemShut
  {NoStop}%
\bibitem [{\citenamefont {Guo}\ \emph {et~al.}(2018)\citenamefont {Guo},
  \citenamefont {Hanhart}, \citenamefont {Mei\ss{}ner}, \citenamefont {Wang},
  \citenamefont {Zhao},\ and\ \citenamefont {Zou}}]{Guo:2017jvc}%
  \BibitemOpen
  \bibfield  {author} {\bibinfo {author} {\bibfnamefont {F.-K.}\ \bibnamefont
  {Guo}}, \bibinfo {author} {\bibfnamefont {C.}~\bibnamefont {Hanhart}},
  \bibinfo {author} {\bibfnamefont {U.-G.}\ \bibnamefont {Mei\ss{}ner}},
  \bibinfo {author} {\bibfnamefont {Q.}~\bibnamefont {Wang}}, \bibinfo {author}
  {\bibfnamefont {Q.}~\bibnamefont {Zhao}}, \ and\ \bibinfo {author}
  {\bibfnamefont {B.-S.}\ \bibnamefont {Zou}},\ }\href {\doibase
  10.1103/RevModPhys.90.015004} {\bibfield  {journal} {\bibinfo  {journal}
  {Rev. Mod. Phys.}\ }\textbf {\bibinfo {volume} {90}} (\bibinfo {year}
  {2018}),\ 10.1103/RevModPhys.90.015004}\BibitemShut {NoStop}%
\bibitem [{\citenamefont {Ali}\ \emph {et~al.}(2017)\citenamefont {Ali},
  \citenamefont {Lange},\ and\ \citenamefont {Stone}}]{Ali:2017jda}%
  \BibitemOpen
  \bibfield  {author} {\bibinfo {author} {\bibfnamefont {A.}~\bibnamefont
  {Ali}}, \bibinfo {author} {\bibfnamefont {J.~S.}\ \bibnamefont {Lange}}, \
  and\ \bibinfo {author} {\bibfnamefont {S.}~\bibnamefont {Stone}},\ }\href
  {\doibase 10.1016/j.ppnp.2017.08.003} {\bibfield  {journal} {\bibinfo
  {journal} {Prog. Part. Nucl. Phys.}\ }\textbf {\bibinfo {volume} {97}},\
  \bibinfo {pages} {123} (\bibinfo {year} {2017})},\ \Eprint
  {http://arxiv.org/abs/1706.00610} {arXiv:1706.00610 [hep-ph]} \BibitemShut
  {NoStop}%
\bibitem [{\citenamefont {Liu}\ \emph {et~al.}(2019)\citenamefont {Liu},
  \citenamefont {Chen}, \citenamefont {Chen}, \citenamefont {Liu},\ and\
  \citenamefont {Zhu}}]{Liu:2019zoy}%
  \BibitemOpen
  \bibfield  {author} {\bibinfo {author} {\bibfnamefont {Y.-R.}\ \bibnamefont
  {Liu}}, \bibinfo {author} {\bibfnamefont {H.-X.}\ \bibnamefont {Chen}},
  \bibinfo {author} {\bibfnamefont {W.}~\bibnamefont {Chen}}, \bibinfo {author}
  {\bibfnamefont {X.}~\bibnamefont {Liu}}, \ and\ \bibinfo {author}
  {\bibfnamefont {S.-L.}\ \bibnamefont {Zhu}},\ }\href {\doibase
  10.1016/j.ppnp.2019.04.003} {\bibfield  {journal} {\bibinfo  {journal} {Prog.
  Part. Nucl. Phys.}\ }\textbf {\bibinfo {volume} {107}},\ \bibinfo {pages}
  {237} (\bibinfo {year} {2019})},\ \Eprint {http://arxiv.org/abs/1903.11976}
  {arXiv:1903.11976 [hep-ph]} \BibitemShut {NoStop}%
\bibitem [{\citenamefont {Brambilla~et al.}(2020)}]{Brambilla:2019esw}%
  \BibitemOpen
  \bibfield  {author} {\bibinfo {author} {\bibfnamefont {N.}~\bibnamefont
  {Brambilla~et al.}},\ }\href@noop {} {\bibfield  {journal} {\bibinfo
  {journal} {Phys. Rept.}\ }\textbf {\bibinfo {volume} {873}},\ \bibinfo
  {pages} {1} (\bibinfo {year} {2020})},\ \Eprint
  {http://arxiv.org/abs/1907.07583} {arXiv:1907.07583 [hep-ex]} \BibitemShut
  {NoStop}%
\bibitem [{\citenamefont {Kvinikhidze}\ and\ \citenamefont
  {Blankleider}(2021)}]{Kvinikhidze:2021kzu}%
  \BibitemOpen
  \bibfield  {author} {\bibinfo {author} {\bibfnamefont {A.~N.}\ \bibnamefont
  {Kvinikhidze}}\ and\ \bibinfo {author} {\bibfnamefont {B.}~\bibnamefont
  {Blankleider}},\ }\href@noop {} {\  (\bibinfo {year} {2021})},\ \Eprint
  {http://arxiv.org/abs/2102.09558} {arXiv:2102.09558 [hep-th]} \BibitemShut
  {NoStop}%
\end{thebibliography}%

\end{document}